\documentclass[%
 reprint,
 amsmath,amssymb,
 aps,
]{revtex4-2}

\usepackage{graphicx,color}
\usepackage{dcolumn}
\usepackage{bm}
\usepackage{amsmath}
\usepackage{amsfonts}
\usepackage{amssymb}

\begin{document}


\title{Enhanced polariton interaction in the presence of disorder}

\author{Matthew Prest$^{1,2}$}%
\author{Cassandra Imperato$^{1}$}%
\author{Oleg L. Berman$^{1,3}$}%
\author{David W. Snoke$^4$}%
\author{Klaus Ziegler$^{3,5}$}%
\affiliation{%
$^1$The Graduate School and University Center, The City University of New York,
New York, NY 10016, USA\\
$^2$Department of Physics and Astronomy, Hunter College of the City University of New York, 695 Park Avenue,
New York, NY 10065, USA\\
$^3$Physics Department, New York City College of Technology, The City University of New York,
Brooklyn, NY 11201, USA\\
$^4$Department of Physics and Astronomy, University of Pittsburgh, 3941 O’Hara Street, Pittsburgh,
Pennsylvania 15260, USA\\
$^5$Institut f\"ur Physik, Universit\"at Augsburg, D-86135 Augsburg, Germany}


\date{\today}

\begin{abstract}
We consider the interaction between exciton-polaritons in a semiconductor quantum well,
embedded in a microcavity, in the presence of disorder. The disorder acts on the excitons  
in the semiconductor quantum well. We have calculated the exciton and polariton 
self-energies and the exciton and polariton energy dispersion relations in the 
presence of disorder. Our results demonstrate that disorder increases the polariton-polariton
interaction.
\end{abstract}

\maketitle


\section{\label{intro} Introduction}
Absorption of a photon by a semiconductor leads to the creation of an electron in the conduction band and a positive charge, i.e., "hole", in the valence band. This electron-hole pair can form a bound state called an "exciton" \cite{snoke2020solid, Moskalenko_Snoke}. The Bose-Einstein condensation and superfluidity of such excitons are expected to exist at experimentally observed exciton densities at temperatures much higher than for the BEC of alkali atoms \cite{Moskalenko_Snoke}. A direct exciton is a two-dimensional (2D) exciton, formed as a bound state by an electron and a hole in a single semiconductor quantum well, while an indirect exciton is formed by the bound state of an electron and a hole in neighboring quantum wells. Excitons can be created when the material absorbs photons and can decay by emitting photons. When a suitable material for the occurrence of excitons is put inside an optical microcavity, linear superposition between photons and excitons can be found \cite{snoke2017new}. Such a quasi-particle is known as an exciton-polariton.

Many theoretical and experimental studies have identified Bose coherent effects of 2D excitonic polaritons in a quantum well embedded in a semiconductor microcavity \cite{littlewood2007condensates, carusotto2013quantum, snoke2020solid, snoke2017new}. To obtain polaritons, two Bragg mirrors are placed opposite each other in order to form a microcavity, and a quantum well is embedded within the cavity at the antinodes of the confined optical mode. The resonant interaction between a direct exciton in a quantum well and a microcavity photon results in the Rabi splitting of the excitation spectrum. Two polariton branches appear in the spectrum due to the resonant exciton-photon coupling. The lower polariton branch of the spectrum has a minimum at zero momentum. The effective mass of the lower polariton is extremely small. These lower polaritons form a 2D weakly interacting Bose gas. The extremely light mass of these bosonic quasiparticles at experimentally achievable excitonic densities results in a relatively high critical temperature for superfluidity. The critical temperature is relatively high because the 2D thermal de Broglie wavelength is inversely proportional to the mass of the quasiparticle, and this wavelength becomes comparable to the distance between the bosons. BEC and superfluidity of exciton-polaritons have been observed in a microcavity \cite{carusotto2013quantum, snoke2020solid, peng2022room}. The various applications of microcavity polaritons for optoelectronics and nanophotonics have recently been developed \cite{snoke2017new}.

The influence of disorder on polariton-polariton interaction was studied experimentally in Ref. \cite{snoke2023reanalysis}. In this paper we develop the theory of polariton-polariton interaction in the presence of disorder, acting on the excitons. We have obtained the exciton and polariton self-energies and exciton and polariton dispersion relations in the presence of disorder,acting on the excitons. We have obtained the exciton self-energy in the coherent potential approximation (CPA). We have obtained the polariton-polariton interaction by employing the Lindhard screening model for a Bose gas \cite{snoke2023reanalysis}. We have demonstrated that the polariton-polariton interaction is enhanced due to disorder.
\section{Disordered Excitons}
We consider a polariton gas in the presence of disorder, where the electron is interacting with a potential $U_e = \alpha_e U(\vec{r})$, and a hole is interacting with the potential $U_h = -\alpha_h U(\vec{r})$. Disorder is presented by an uncorrelated white noise potential $U(\vec{r})$ with
\begin{align}\label{white noise}
    \langle U(\vec{r}) U(\vec{r}`) \rangle = D \delta^{d}(\vec{r} - \vec{r}`) \text{, and } \langle U(\vec{r}) \rangle = 0.
\end{align}

\begin{figure}[t]
    \centering
   \includegraphics[width=1.1\linewidth]{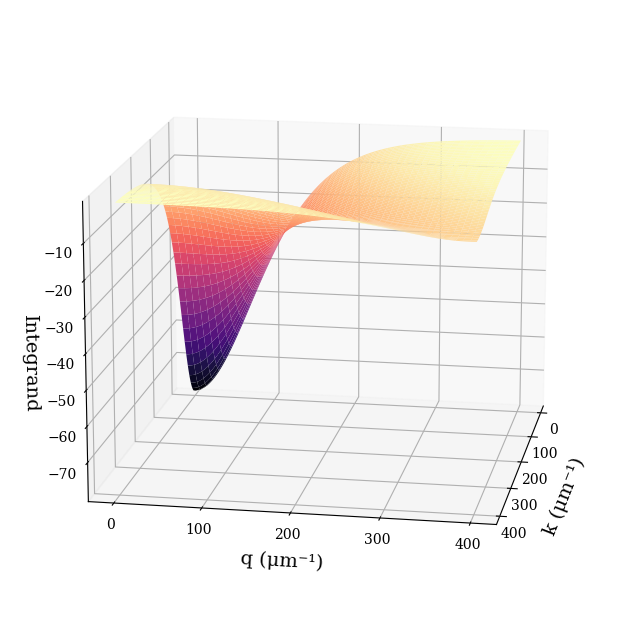}
    \caption{A surface plot of the integrand weight $K(k,q)$ for $m'=1/3$ over a domain of $k$ and $q$ up to $4 \times 10^5 cm^{-1}$, illustrating convergence to $0$ for large values of $q$.}
    \label{Integrand Weight}
\end{figure}

A derivation of the self-energy $Q(\vec{k})$ of the polariton in $d=2$ with disorder strength $D$, and $m_e \neq m_h$ can be based on the self-consistent CPA,
given by the CPA equation~\cite{gevorkyan1985localization}
\begin{align}
    Q(\vec{k}) = \frac{-1}{2} \int G^0(\vec{q}) B(|\vec{k} - \vec{q}|) \frac{d^2 q}{4\pi^2},
\end{align}

where $G^0$ is the associated Green's function of the exciton gas. Thus, we get from the proper choice of $G^0$ and $B$ the equation (cf. discussion in Ref.\cite{gevorkyan1985localization})
\begin{align}
    Q(\vec{k}) &= \int \frac{d^2q}{E - \frac{\hbar^2q^2}{2M} + Q(q)+i\epsilon} \bigg[ \frac{\alpha_e}{m_h} \big(|\vec{k} - \vec{q}|^2 + \nonumber \\
     & \frac{4M^2}{a^2 m_e^2} \big)^{-3/2} - \frac{\alpha_h}{m_e} \big(|\vec{k} - \vec{q}|^2 + \frac{4M^2}{a^2 m_h^2} \big)^{-3/2} \bigg]^2.
\end{align}
where $\epsilon>0$ was introduced to avoid the poles of the
Green's function.
First we note that this equation is rotational invariant, which reduces the CPA equation to a
$1D$ integral equation of the form
\begin{align}
\label{hammerstein}
    Q(k) &= -\frac{2DM^4}{m_e m_h \pi a^4} \int_0^{\infty} \frac{q dq}{E - \frac{\hbar^2 q^2}{2M} + Q(q) + i\epsilon}  \nonumber \\
    & \times \int_0^{2\pi} d\theta \bigg[ \frac{\alpha_e}{m_h}\bigg( |\vec{k}-\vec{q}|^2 + \frac{4M^2}{a^2 m_e^2}\bigg)^{-3/2} \nonumber \\
    & - \frac{\alpha_h}{m_e}\bigg( |\vec{k}-\vec{q}|^2 + \frac{4M^2}{a^2 m_h^2}\bigg)^{-3/2} \bigg]^2,
\end{align}
where $\alpha_e$, $\alpha_h$ are inversely proportional to their respective masses
\begin{align}
    \alpha_e = \frac{m' m_0}{m_e}
    , \ \
    \alpha_h = \frac{m' m_0}{m_h}
\end{align}
with $m_0$ the rest mass of the electron in vacuum and $m'$ is a tunable parameter. The type of equation (\ref{hammerstein}) is also known as a non-linear Hammerstein Equation \cite{dolph1949nonlinear} with the general form
\begin{align}
    \varphi(x)=\int_{a}^{b} K(x, s) F(s, \varphi(s)) ds.
\end{align}
For a numerical solution of this equation we employ the Picard Iteration approach~\cite{kelley1995iterative}
with a damping parameter $\omega$:
\begin{align}
    Q_{n+1}(k)= (1-\omega)Q_n(k) + \omega\int K(q, k) \cdot F(q, Q_{n}(q)) dq.
\end{align}
To examine the feasibility of solving this equation numerically, we plot the integrand weight $K(k,q)$ in
Fig. \ref{Integrand Weight}. As illustrated, the most significant contribution that will affect $Q(k)$
occurs near the origin, and a large, finite cutoff for $q$ will be sufficient.

\section{Lower Polariton Dispersion}

Polaritons emerge through the coupling of the excitons to cavity photons. This results in a linear superposition of excitons with the cavity photons. Here we assume that disorder does not affect
the photons. Therefore, we consider disordered excitons, following the results of the previous section, and couple them directly to the photons, which leads to a linear superposition of disordered excitons and photons. Then the dispersion relation for the lower polariton \cite{snoke2023reanalysis} is given by
\begin{align}\label{E_LP}
    E_{LP}(\vec{k}) &= \frac{1}{2}(E_{ph}(\vec{k}) + E_{ex}(\vec{k}) - \nonumber \\
    & \sqrt{[E_{ph}(\vec{k}) - E_{ex}(\vec{k})]^2 + 4\Omega^2}),
\end{align}
where $\Omega$ is the Rabi splitting and
\begin{align}
    E_{ex}(\vec{k}) = E_{band} - E_{bind} + \frac{\hbar^2k^2}{2M} - Q({\vec{k}}).
\end{align}
Additionally, the photon energy is determined by
\begin{align}
    E_{ph} = \sqrt{\bigg(\frac{\hbar c k}{n}\bigg)^2 + k_0^2},
\end{align}
where $n$ is the cavity refractive index and $k_0$ is set to the exciton energy at $k=0$ such that the detuning is zero.
\section{Polariton-Polariton Interaction Strength}
To calculate the effective interaction strength between the polaritons at finite temperature we use the Matsubara formalism that was outlined in \cite{snoke2020solid}. To this end, we first calculate the Hopfield coefficients \cite{hopfield1958theory}
\begin{align}
    X_P = \frac{1}{\sqrt{1 + \big(\frac{\Omega}{E_{LP} - E_{ph}} \big)^2}}\\
    C_P = \frac{-1}{\sqrt{1 + \big(\frac{E_{LP} - E_{ph}}{\Omega} \big)^2}}.
\end{align}
%
Next, the mean-field polariton-polariton interaction strength $g$
without correlations can be written as~\cite{snoke2023reanalysis}
\begin{align}
\label{g_bare}
    g = g_{ex}|X_P|^4,
\end{align}
where we use $g_{ex} = 12 \hspace{1mm} \mu \text{eV}\mu \text{m}^2$ \cite{snoke2023reanalysis}. Finally, we calculate the polarization bubble $\Pi^0$, using the Matsubara finite-temperature formalism for 2D bosons \cite{snoke2020solid, snoke2023reanalysis, fetter2012quantum}
\begin{align}
    \Pi^0 = -\frac{A M_{\text{pol}}}{\pi \hbar^2} \frac{1}{e^{-\mu/k_B T} -1},
\end{align}
where $M_{\mathrm{pol}}$ is the effective polariton mass. This is
plugged into the polariton-polariton interaction strength taking
into account the correlation effects $g^{\prime}$ following the
procedure in Ref.\cite{snoke2023reanalysis} 
and Ref.\cite{fetter2012quantum} to give
\begin{align}
\label{g_prime}
    g' \approx g \frac{1 - \frac{g}{A}\Pi^0}{1 - 2\frac{g}{A}\Pi^0}.
\end{align}
This approximation holds for $\frac{g}{A}\Pi^0 << 1/2$ which is valid in the low density, non-condensate regime that we examine here.
%
In this case, the polariton-polariton interaction strength
with the correlation effects $g^{\prime}$ given by Eq.~(\ref{g_prime}) will be
approximately 5\% larger than the  mean-field polariton-polariton
interaction strength $g$ given by Eq.~(\ref{g_bare}).
\section{Experimental Values}
\begin{figure}[t]
    \centering
    \includegraphics[width=\linewidth]{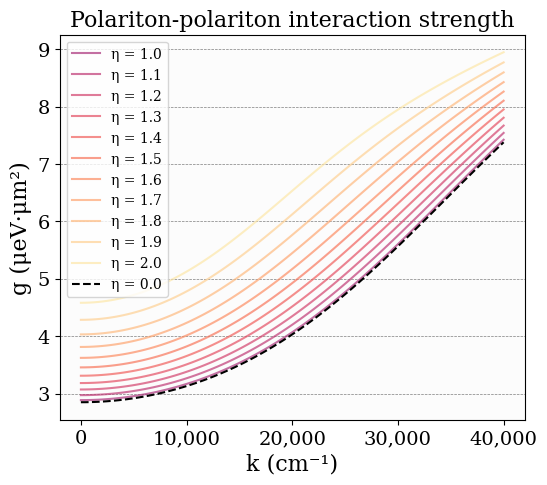}
    \caption{The mean field polariton-polariton
    interaction strength $g$.
    }
    \label{interaction_strength}
\end{figure}
\begin{figure}[t]
    \centering
    \includegraphics[width=\linewidth]{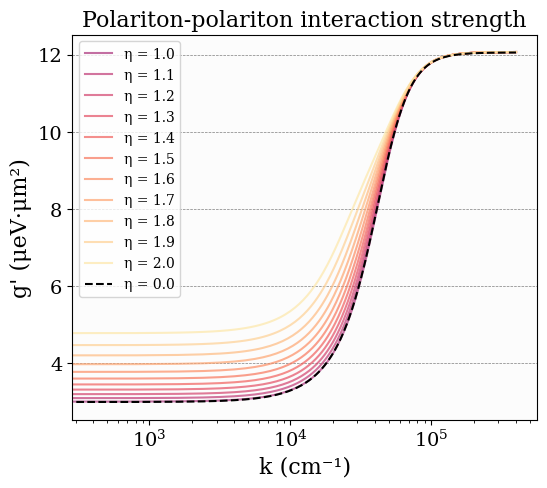}
    \caption{The polariton-polariton interaction strength with correlation effects $g^{\prime}$ for a range of disorder factors $\eta \in [0.0, 2.0]$ over momentum space
    with a logarithmic $k$-scale with $m'=1/3$.}
    \label{interaction_strength_k_log}
\end{figure}
To derive the experimental value for disorder strength, we first relax the assumed delta correlation function to a Gaussian,
\begin{align}\label{exp corr}
    \langle U(\vec{r}) \cdot U(\vec{r'}) \rangle = \Delta^2 e^{-|\vec{r} - \vec{r'}|^2/2\sigma^2}
\end{align}
where $\Delta$ is the energy fluctuation and $\sigma$ is the correlation length. We can then integrate Eqs. (\ref{exp corr}) and (\ref{white noise}) over all space and equate the results.
\begin{align}
    \int D \delta^2(\vec{r} - \vec{r'}) d^2 r' = \int \Delta^2 e^{-|\vec{r} - \vec{r'}|^2/2\sigma^2} d^2 r'.
\end{align}
This gives the result of
\begin{align}
    D = 2 \pi \Delta^2 \sigma^2.
\end{align}
With an energy fluctuation $\Delta = 3 \text{ meV}$, and correlation length $\sigma = 20 \text{ nm}$ we have a value of $D_0 = 2.26 \times 10^{-20} eV^2 m^2$.
To adjust this result with realistic values for the disorder strength, we then explore a range of disorder factors $\eta$, where $D = \eta D_0$. Moreover, for 2D Gallium Arsenide we consider the following energies: $E_{band} = 1.6$ eV, $E_{bind} = 4.2$ meV, and for the Rabi Splitting $\Omega = 14 \text{ meV}$. We used a refractive index inside the cavity of $n=3.0$.
The effective mass of the polariton for a slowly changing confinement potential can be derived from the curvature of the band minimum in the manner of Ref. \cite{berman2008theory}. As our dispersion is isotropic we use
\begin{align}\label{m_eff}
    \frac{1}{M_{\text{pol}}} = \frac{1}{\hbar^2} \Bigg[\frac{d^2 Re(E_{LP})}{dk^2}\Bigg]_{k=0}.
\end{align}
After inserting (\ref{E_LP}) into (\ref{m_eff}) we get a narrow range of disorder-dependent effective masses that are on the scale of $10^{-5}m_0$.
For the experimental temperature of the polaritons we consider $20 K$ with the polariton concentration $n = 0.3 \hspace{1mm} \mu m^{-2}$ \cite{snoke2023reanalysis}. We use this to solve for the disorder-dependent chemical potential $\mu (\eta)$ using a fugacity series approach \cite{pathria2017statistical}.
Because our lower polariton dispersion is isotropic and using the particle number $N = n A$ for a given area, we begin with the Boson density of states integral relation
\begin{align}
    n A = \frac{A}{2\pi} Re\int_0^{\infty} \frac{k \hspace{1mm} dk}{e^{\beta(E_{LP}(k) -\mu)} -1}.
\end{align}
We convert this to a series in terms of the fugacity $z \equiv e^{\beta \mu}$ to get
\begin{align}
    n = \frac{1}{2\pi} \sum_{l=1}^{\infty} z^l Re\int_0^{k_{max}} k e^{-l\beta E_{LP}(k)} dk.
\end{align}
To estimate this numerically, we truncate the integral to $k_{\text{max}} = 10^6 \text{cm}^{-1}$, well above experimental ranges. The sum converged rapidly and likewise was truncated at $l = 100$.
%
%
\section{Results}
\begin{figure*}
    \centering
    \includegraphics[width=\linewidth]{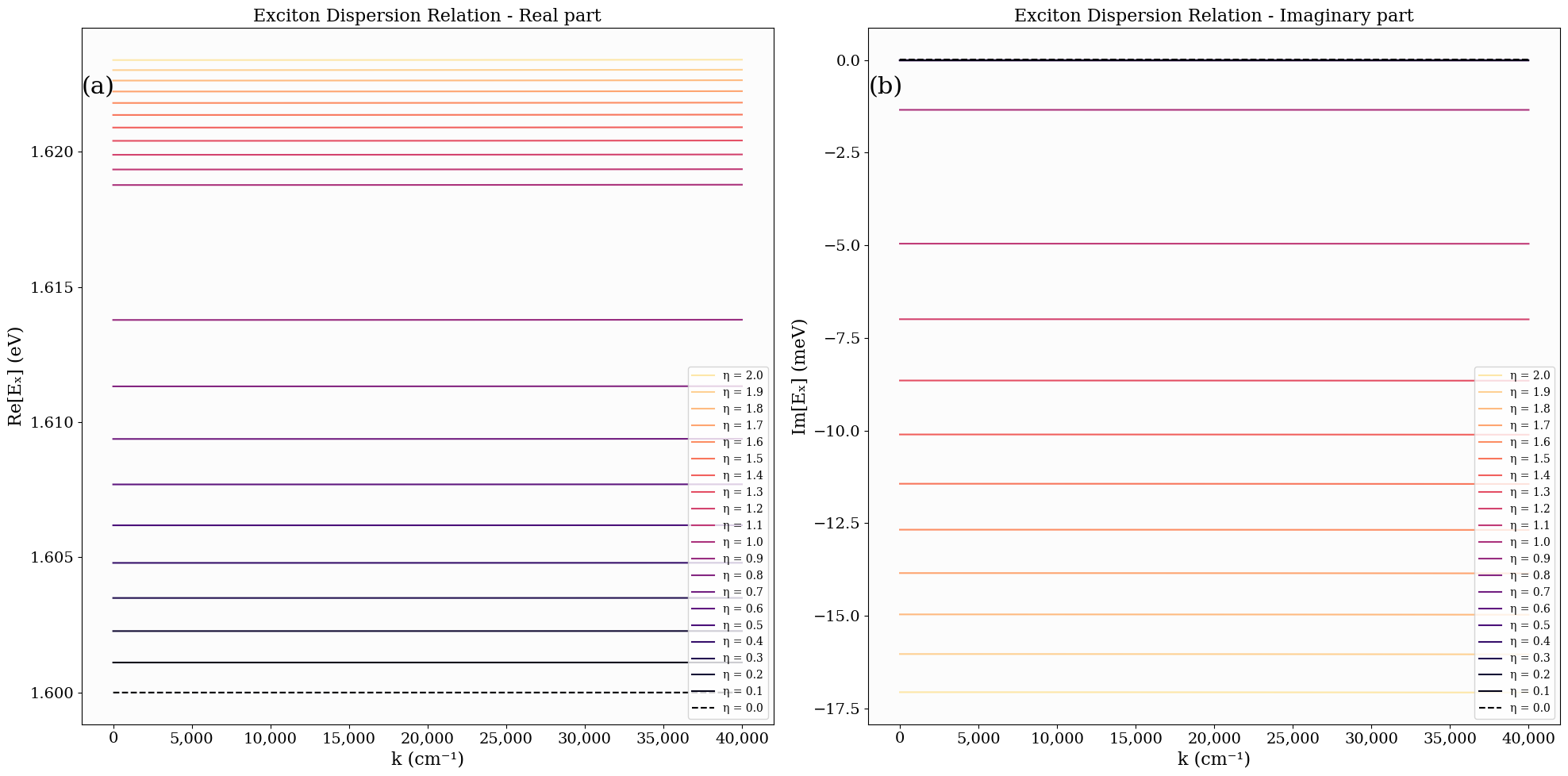}
    \caption{The real (a) and imaginary (b) parts of the exciton energy dispersion relation in the presence of disorder  characterized by the range $\eta \in [0.0, 2.0]$ over momentum space with $m'=1/3$.
    }
    \label{exciton_dispersion}
\end{figure*}
\begin{figure*}
    \centering
    \includegraphics[width=\linewidth]{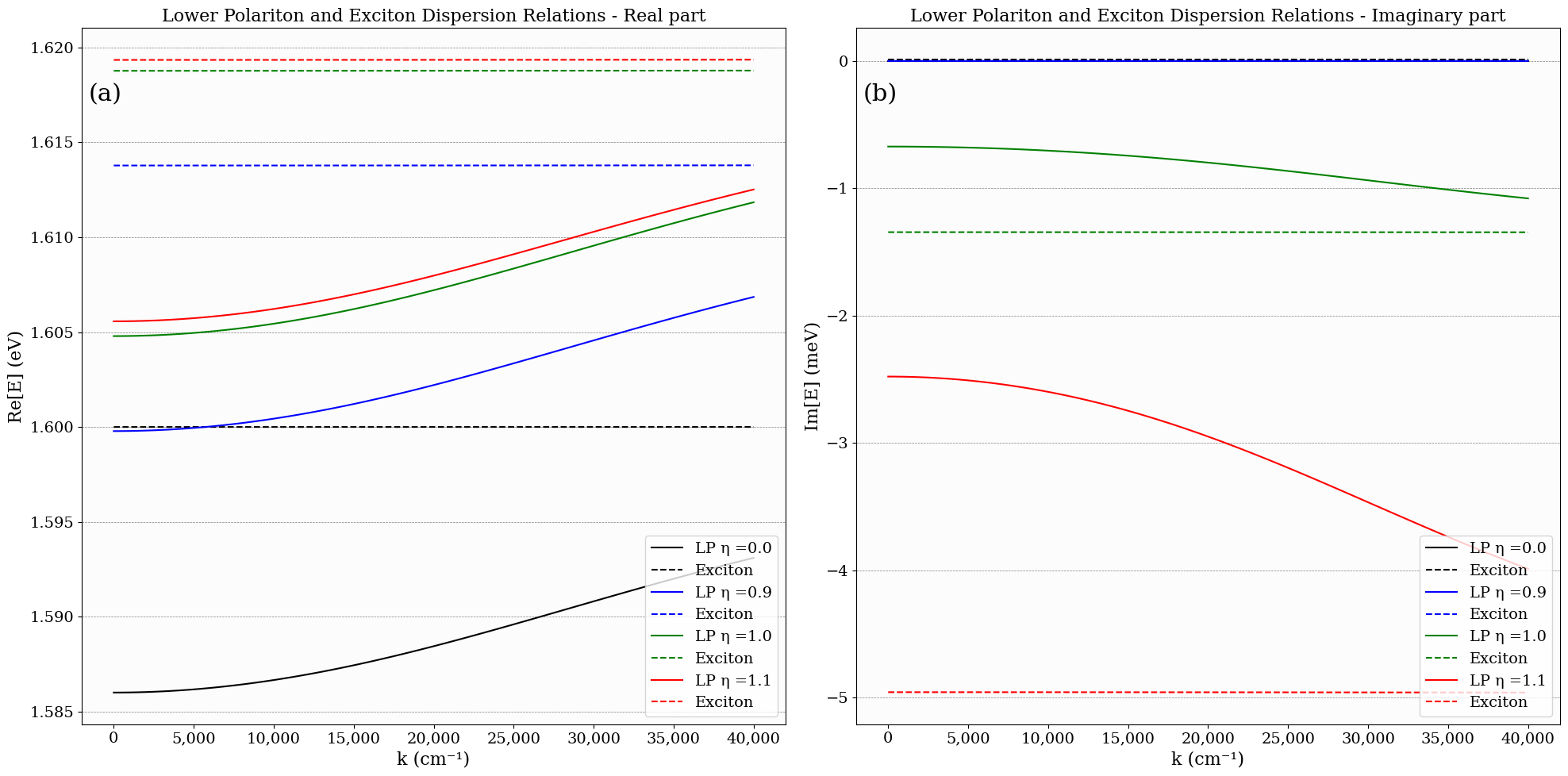}
    \caption{The real (a) and imaginary (b) parts of the lower polariton energy dispersion relation in the presence of disorder  characterized by the values $\eta \in [0.0, 0.9, 1.0, 1.1]$ over momentum space with $m'=1/3$.}
    \label{polariton_dispersion}
\end{figure*}
We have demonstrated in Fig. \ref{interaction_strength} that the
polariton-polariton interaction for resonant lower polaritons is
enhanced due to disorder, acting on excitons. The expected behavior,
of monotonically increasing $g'$ with larger $k$, corresponding to
the polariton becoming more excitonic is demonstrated. Additionally,
there is a clear positive shift in interaction strength when
disorder is increased across the range of $k$. As it can be seen in 
Fig.~\ref{interaction_strength_k_log}, 
the polariton-polariton
interaction strength $g^{\prime}$ at high $k$ saturates to the
constant of exciton-exciton interaction $g_{ex}$. This is due to the
fact that in Eq.~(\ref{g_bare}) the Hopfield coefficient $X_{P}$
asymptotically goes to $1$ at large $k$. In
Fig.~\ref{exciton_dispersion} we see the effects of our numerically
derived self-energy on the exciton dispersion relation. We see both
an increase in exciton energy for larger disorder and a more
negative imaginary part of the exciton dispersion corresponding to
increased line broadening and reduced particle lifetime in the
presence of stronger disorder as expected. The spacing between lines
of differing disorder shows a non-linear dependence on the disorder
factor $\eta$. The increase in the real part of the energy increases
up to the largest jump from $\eta=0.9$ to $\eta=1.0$ where the
difference then proceeds to decrease for larger $\eta$ possibly
indicating some sort of disorder saturation. The imaginary component
also exhibits a similar jump for the same values of $\eta$. The
value $m'=1/3$ was selected so that the experimental estimate of
disorder corresponds $\eta=1.0$ to a broadening of order
$1\text{meV}$ \cite{snoke2023reanalysis}.
In the final figure, Fig. \ref{polariton_dispersion} the lower polariton dispersion relation is depicted in a narrow range of values of $\eta$ showing the convergence to exciton energy for large $k$ and the impact of disorder as a positive shift to the real part of the energy and a negative shift to the imaginary part corresponding to a reduced polariton lifetime. The behavior demonstrated is the same as for the exciton, where Fig. \ref{polariton_dispersion} highlights the largest jump in the dispersion relation of the polariton.
\section{Conclusion}
Our calculations are the first to our knowledge to include disorder as a self-energy term in the exciton dispersion, and indicate an increasing interaction strength between polaritons in the presence of disorder. While this increase nominally contributes to a closer predicted value to experiment, it remains substantially lower due to the input value of $g_{ex}$. Future consideration possibly including multiple interacting quantum wells may be needed to better replicate experimental results.
\section{Acknowledgments}
The work of D.W.S. has been supported by the NSF Grant DMR-2306977.
\section{Appendix}
We present the full derivation of the self-energy of the polariton in $d=2$ with disorder strength $D$, and $m_e \neq m_h$. Recall the spatial correlation function \cite{gevorkyan1985localization}:
\begin{align}
    & B(\vec{R}_{1}, \vec{R_2})=D \int d^d r \bigg[ \alpha_{e} \bigg(\frac{M}{m_h}\bigg)^d \bigg|\varphi_{0}\bigg((\vec{r}-\vec{R_{1}}) \frac{M}{m_{h}}\bigg)\bigg|^{2} \nonumber \\
    & -\alpha_{h}\bigg(\frac{M}{m_{e}}\bigg)^{d}\bigg|\varphi_{0}\bigg((\vec{R_{1}}-\vec{r}) \frac{M}{m_{e}}\bigg)\bigg|^{2}\bigg] \cdot \bigg[\bigg.\alpha_{e}\bigg(\frac{M}{m_{h}}\bigg)^{d} \times \nonumber \\
    & \bigg|\varphi_{0}\bigg((\vec{r}-\vec{R}_{2}) \frac{m}{m_{h}}\bigg|^{2}-\alpha_{h}\bigg(\frac{m}{m_{e}}\bigg)^{d} \bigg\rvert\, \varphi_{0}\bigg((\vec{R}_{2}-\vec{r}) \frac{M}{m_{e}}\bigg)\bigg|^{2}\bigg]. \nonumber \\
\end{align}
Taking a translation such that $\vec{R}_{2}=\vec{0}$ and setting $d=2$ implies,
\begin{align}
    &B(\vec{R_1}) = D \int d^{2} r\bigg[\alpha_{e}\bigg(\frac{M}{m_{h}}\bigg)^{2}\bigg|\varphi_{0}\bigg(\frac{M}{m_{h}}(\vec{r}-\vec{R_{1}})\bigg)\bigg|^{2} - \nonumber \\
    & \bigg.\alpha_{h}\bigg(\frac{M}{m_{e}}\bigg)^{2}\bigg|\varphi_{0}\bigg((\vec{R_{1}}-\vec{r}) \frac{M}{m_{e}}\bigg)\bigg|^{2}\bigg] \times \bigg[\alpha_{e}\bigg(\frac{M}{m_{h}}\bigg)^{2}\bigg|\varphi_{0}(\vec{r} \frac{M}{m_{h}})\bigg|^{2} \nonumber \\
    & -\alpha_{h}\bigg(\frac{M}{m_{e}}\bigg)^{2}\bigg|\varphi_{0}(\vec{r} \frac{M}{m_{e}})\bigg|^{2}\bigg].
\end{align}
Using the $2D$ ground state wave function
\begin{align}
    |\varphi_0(r)|^2 = e^{-2r/a}\pi a^2,
\end{align}
and defining the following constants, $\beta_h = 2M/am_h$ and $\beta_e = 2M/am_e$, we arrive at the following form,
\begin{align}
    & B(\vec{R_1}) = \frac{DM^2}{\pi a^2} \int d^{2} r\bigg[ \frac{\alpha_e^2}{m_h^4} e^{-\beta_h|\vec{r}-\vec{R_1}|} e^{-\beta_h r} -  \nonumber \\
    & \frac{\alpha_e \alpha_h}{m_h^2 m_e^2} e^{-\beta_h|\vec{r}-\vec{R_1}|} e^{-\beta_e r} - \frac{\alpha_e \alpha_h}{m_h^2 m_e^2} e^{-\beta_e|\vec{r}-\vec{R_1}|} e^{-\beta_h r} + \nonumber \\
    & \frac{\alpha_h^2}{m_e^4} e^{-\beta_e|\vec{r}-\vec{R_1}|} e^{-\beta_e r}\bigg].
\end{align}
We then can break apart $B(\vec{R_1})$ into the following four integrals,
\begin{align}
    & I_1 = \frac{DM^2\alpha_e^2}{\pi a^2 m_h^4} \int d^2r \hspace{1mm} e^{-\beta_e r} e^{-\beta_e |\vec{r} - \vec{R_1}|}, \\
    & I_2 = -\frac{DM^2\alpha_e \alpha_h}{\pi a^2 m_e^2 m_h^2} \int d^2r \hspace{1mm} e^{-\beta_e r} e^{-\beta_h |\vec{r} - \vec{R_1}|}, \\
    & I_3 = -\frac{DM^2\alpha_e \alpha_h}{\pi a^2 m_e^2 m_h^2} \int d^2r \hspace{1mm} e^{-\beta_h r} e^{-\beta_e |\vec{r} - \vec{R_1}|}, \\
    & I_4 = \frac{DM^2\alpha_h^2}{\pi a^2 m_e^4} \int d^2r \hspace{1mm} e^{-\beta_h r} e^{-\beta_h |\vec{r} - \vec{R_1}|}.
\end{align}
We can define the following functions so that each $I_i$ can be expressed as a convolution, $f_e(\vec{r}) = e^{-\beta_e r}$ and $f_h(\vec{r}) = e^{-\beta_h r}$. Our integrals can now be expressed as follows
\begin{align}
    & I_1 =  \frac{DM^2\alpha_e^2}{\pi a^2 m_h^4} \big[ f_e * f_e \big] (\vec{R_1}), \\
    & I_2 =  -\frac{DM^2\alpha_e \alpha_h}{\pi a^2 m_e^2 m_h^2} \big[ f_e * f_h \big] (\vec{R_1}), \\
    & I_3 =  -\frac{DM^2\alpha_e \alpha_h}{\pi a^2 m_e^2 m_h^2} \big[ f_h * f_e \big] (\vec{R_1}), \\
    & I_4 =  \frac{DM^2\alpha_h^2}{\pi a^2 m_e^4} \big[ f_h * f_h \big] (\vec{R_1}).
\end{align}
Now taking the Fourier transforms of $f_e$ and $f_h$
\begin{align}
    \Tilde{f}_e(k) &= \int e^{i\vec{k}.\vec{r}} e^{-\beta_e r} r dr d\theta \nonumber \\
    & = \int_0^{\infty} r e^{-\beta_e r} dr \int_0^{2\pi} e^{ikr\cos\theta} d\theta \nonumber \\
    & = \int_0^{\infty} r e^{-\beta_r} 2\pi J_0(kr) dr,
\end{align}
where $J_0(kr)$ is the Bessel Function of the first kind. Now note that for a Laplace transform with $s > 0$,
\begin{align}
    \mathcal{L} [J_0(kr)] = \int_0^{\infty} e^{-sr} J_0(kr) dr = \frac{1}{\sqrt{s^2 + k^2}}.
\end{align}
We can introduce a derivative operator in $s$,
\begin{align}
    \frac{d}{ds} \bigg[\int_0^{\infty} e^{-sr} J_0(kr) dr\bigg] &= \int_0^{\infty} \frac{d}{ds} \bigg[ e^{-sr}\bigg ] J_0(kr) dr \nonumber \\
    & = - \int_0^{\infty} r e^{-sr}  J_0(kr) dr \nonumber \\
    & = \frac{d}{ds} \frac{1}{\sqrt{s^2 + k^2}}.
\end{align}
Calculating the derivative and substituting back in for $s$,
\begin{align}
    \implies \Tilde{f}_e(k) = \frac{2\pi\beta_e}{(k^2 + \beta_e^2)^{3/2}}, \Tilde{f}_h(k) = \frac{2\pi\beta_h}{(k^2 + \beta_e^2)^{3/2}}.
\end{align}
Using that the convolution is equal to the inverse Fourier transform of the product in Fourier space gives us
\begin{align}
    B(\vec{k}) = \frac{16 \pi D M^4}{m_e m_h a^4} \bigg[&\frac{\alpha_e}{m_h} (k^2 + 4M^2/a^2m_e^2)^{-3/2} - \nonumber \\
    &\frac{\alpha_h}{m_e} (k^2 + 4M^2/a^2m_h^2)^{-3/2}\bigg]^2.
\end{align}
Which gives us
\begin{align}
    B(|\vec{k} - \vec{q}|) &=  \frac{16 \pi D M^4}{m_e m_h a^4} \bigg[\frac{\alpha_e}{m_h} (|\vec{k} - \vec{q}|^2 + 4M^2/a^2m_e^2)^{-3/2}  \nonumber \\
    & - \frac{\alpha_h}{m_e} (|\vec{k} - \vec{q}|^2 + 4M^2/a^2m_h^2)^{-3/2}\bigg]^2.
\end{align}
Recall from Ref. \cite{gevorkyan1985localization}
\begin{align}
    Q(\vec{k}) = \frac{-1}{2} \int G^0(\vec{q}) B(|\vec{k} - \vec{q}|) \frac{d^2 q}{4\pi^2},
\end{align}
Where $G^0$ is the associated Green's function. Substituting in gives us
\begin{align}
    Q(\vec{k}) &= \int \frac{d^2q}{E - \frac{\hbar^2q^2}{2M} + Q(q)} \bigg[ \frac{\alpha_e}{m_h} \big(|\vec{k} - \vec{q}|^2 + \nonumber \\
     & \frac{4M^2}{a^2 m_e^2} \big)^{-3/2} - \frac{\alpha_h}{m_e} \big(|\vec{k} - \vec{q}|^2 + \frac{4M^2}{a^2 m_h^2} \big)^{-3/2} \bigg]^2.
\end{align}
First we can use rotational symmetry to reduce to a $1D$ integral equation of the same form,
\begin{align}
    Q(k) &= -\frac{2DM^4}{m_e m_h \pi a^4} \int_0^{\infty} \frac{q dq}{E - \frac{\hbar^2 q^2}{2M} + Q(q)}  \nonumber \\
    & \times \int_0^{2\pi} d\theta \bigg[ \frac{\alpha_e}{m_h}\bigg( |\vec{k}-\vec{q}|^2 + \frac{4M^2}{a^2 m_e^2}\bigg)^{-3/2} \nonumber \\
    & - \frac{\alpha_h}{m_e}\bigg( |\vec{k}-\vec{q}|^2 + \frac{4M^2}{a^2 m_h^2}\bigg)^{-3/2} \bigg]^2.
\end{align}
Expanding out the terms gives
\begin{align}
    & Q(k) = -\frac{2DM^4}{m_e m_h \pi a^4} \int_0^{\infty} \frac{q dq}{E - \frac{\hbar^2 q^2}{2M} + Q(q)} \times \nonumber \\
    & \bigg[ \frac{\alpha_e^2}{m_h^2} \int_0^{2\pi} \frac{d\theta}{(k^2 + q^2 -2kq\cos\theta + 4M^2/a^2 m_e^2)^3} - \nonumber \\
    & \frac{\alpha_e \alpha_h}{m_e m_h} \int_0^{2\pi} \frac{d\theta}{(k^2 + q^2 -2kq\cos\theta + 4M^2/a^2 m_e^2)^{3/2}} \nonumber \\
    & \times \frac{1}{(k^2 + q^2 -2kq\cos\theta + 4M^2/a^2 m_h^2)^{3/2}} \nonumber \\
    & + \frac{\alpha_h^2}{m_e^2} \int_0^{2\pi} \frac{d\theta}{(k^2 + q^2 -2kq\cos\theta + 4M^2/a^2 m_h^2)^3}\bigg].
\end{align}
We can solve two of the angular integral terms analytically, first defining
\begin{align}
    I_5 = \frac{\alpha_e^2}{m_h^2} \int_0^{2\pi} \frac{d\theta}{(k^2 + q^2 -2kq\cos\theta + 4M^2/a^2 m_e^2)^3}.
\end{align}
Next we can consider the following similar integral
\begin{align}
    I_6 = \int_0^{2\pi} \frac{x}{a - b \cos x}.
\end{align}
We can use a tangent half-angle substitution $t = \tan x/2$ for the case $|a| > |b|$.
\begin{align}
    \implies I_6 = \int_{\infty}^{\infty} \frac{2 dt}{a+b + (a-b)t^2}.
\end{align}
Which evaluates to
\begin{align}
    I_6 = \frac{2\pi}{\sqrt{a^2 - b^2}}.
\end{align}
Notice that
\begin{align}
    \frac{d}{da}\bigg[ \frac{1}{a - b \cos x} \bigg] = \frac{-1}{(a+b\cos x)^2}.
\end{align}
We can use the "Feynman Trick" to switch the order of operators,
\begin{align}
    \implies & \int_0^{2\pi} \frac{dx}{(a + b\cos x)^2} = \frac{-d}{da} \int_0^{2\pi} \frac{dx}{(a + b\cos x)}, \\
    & = \frac{-d}{da} \bigg[\frac{2\pi}{\sqrt{a^2 - b^2}} \bigg].
\end{align}
We can repeat this trick a second time to get
\begin{align}
    I_5 = \frac{\pi[2(k^2+q^2+4M^2/a^2m_e^2)^2 + 4k^2q^2]}{[(k^2+q^2+4M^2/a^2m_e^2)^2 -4k^2q^2]^{5/2}}.
\end{align}
This form can be used to solve two of the three angular integrals analytically. The remaining cross term instead must be approximated numerically, this term is defined as
\begin{align}
    I_7 &= \int_0^{2\pi} \frac{d\theta}{(k^2 + q^2 -2kq\cos\theta + 4M^2/a^2m_e^2)^{3/2}} \nonumber \\
    & \times \frac{1}{(k^2 + q^2 -2kq\cos\theta + 4M^2/a^2m_h^2)^{3/2}}.
\end{align}
Although no analytic solution for $I_7$ can be found, it is straightforward to approximate numerically. Using this term we are left with a nested numerical integration problem,
\begin{align}
    & Q(k) = \frac{-4DM^4}{m_em_ha^4} \int_0^{\infty} \frac{q dq}{E - \hbar^2q^2/2M + Q(q) + i\epsilon} \times \nonumber \\
    & \bigg[ \frac{\alpha_e^2}{m_h^2} \frac{(k^2+q^2+4M^2/a^2m_e^2)^2 + 2k^2q^2}{[(k^2+q^2+4M^2/a^2m_e^2)^2 - 4k^2q^2]^{5/2}} + \frac{\alpha_h^2}{m_e^2} \times\nonumber \\
    &  \frac{(k^2+q^2+4M^2/a^2m_h^2)^2 + 2k^2q^2}{[(k^2+q^2+4M^2/a^2m_h^2)^2 - 4k^2q^2]^{5/2}} - \frac{\alpha_e\alpha_h}{m_e m_h \pi} I_7 \bigg].
\end{align}
With $\alpha_e$, $\alpha_h$ the relative disorder sensitivities, inversely proportional to their respective masses,
\begin{align}
    \alpha_e = \frac{m'}{m_e}, \\
    \alpha_h = \frac{m'}{m_h},
\end{align}
with $m'$ approximately equal to the rest mass of the electron in vacuum. Our Hammerstein equation now becomes
\begin{align}
    Q(k) &= \frac{-4DM^4 m'^2}{m_e^3 m_h^3 a^4} \int_0^\infty \frac{qdq}{E - \hbar^2q^2/2M + Q(q) + i\epsilon} \\ \nonumber
    &\times \bigg[\frac{(k^2+q^2+4M^2/a^2m_e^2)^2 + 2k^2q^2}{[(k^2+q^2+4M^2/a^2m_e^2)^2 - 4k^2q^2]^{5/2}} + \\ \nonumber
    &+ \frac{(k^2+q^2+4M^2/a^2m_h^2)^2 + 2k^2q^2}{[(k^2+q^2+4M^2/a^2m_h^2)^2 - 4k^2q^2]^{5/2}}
    - \frac{1}{\pi} I_7 \bigg].
\end{align}
Our resulting equation is a non-linear Hammerstein Equation \cite{dolph1949nonlinear}, of the form
\begin{align}
    \varphi(x)=\int_{a}^{b} K(x, s) F(s, \varphi(s)) ds.
\end{align}
%


%


\begin{thebibliography}{14}%
\makeatletter
\providecommand \@ifxundefined [1]{%
 \@ifx{#1\undefined}
}%
\providecommand \@ifnum [1]{%
 \ifnum #1\expandafter \@firstoftwo
 \else \expandafter \@secondoftwo
 \fi
}%
\providecommand \@ifx [1]{%
 \ifx #1\expandafter \@firstoftwo
 \else \expandafter \@secondoftwo
 \fi
}%
\providecommand \natexlab [1]{#1}%
\providecommand \enquote  [1]{``#1''}%
\providecommand \bibnamefont  [1]{#1}%
\providecommand \bibfnamefont [1]{#1}%
\providecommand \citenamefont [1]{#1}%
\providecommand \href@noop [0]{\@secondoftwo}%
\providecommand \href [0]{\begingroup \@sanitize@url \@href}%
\providecommand \@href[1]{\@@startlink{#1}\@@href}%
\providecommand \@@href[1]{\endgroup#1\@@endlink}%
\providecommand \@sanitize@url [0]{\catcode `\\12\catcode `\$12\catcode `\&12\catcode `\#12\catcode `\^12\catcode `\_12\catcode `\%12\relax}%
\providecommand \@@startlink[1]{}%
\providecommand \@@endlink[0]{}%
\providecommand \url  [0]{\begingroup\@sanitize@url \@url }%
\providecommand \@url [1]{\endgroup\@href {#1}{\urlprefix }}%
\providecommand \urlprefix  [0]{URL }%
\providecommand \Eprint [0]{\href }%
\providecommand \doibase [0]{https://doi.org/}%
\providecommand \selectlanguage [0]{\@gobble}%
\providecommand \bibinfo  [0]{\@secondoftwo}%
\providecommand \bibfield  [0]{\@secondoftwo}%
\providecommand \translation [1]{[#1]}%
\providecommand \BibitemOpen [0]{}%
\providecommand \bibitemStop [0]{}%
\providecommand \bibitemNoStop [0]{.\EOS\space}%
\providecommand \EOS [0]{\spacefactor3000\relax}%
\providecommand \BibitemShut  [1]{\csname bibitem#1\endcsname}%
\let\auto@bib@innerbib\@empty
\bibitem [{\citenamefont {Snoke}(2020)}]{snoke2020solid}%
  \BibitemOpen
  \bibfield  {author} {\bibinfo {author} {\bibfnamefont {D.~W.}\ \bibnamefont {Snoke}},\ }\href@noop {} {\emph {\bibinfo {title} {Solid state physics: Essential concepts}}}\ (\bibinfo  {publisher} {Cambridge University Press},\ \bibinfo {year} {2020})\BibitemShut {NoStop}%
\bibitem [{\citenamefont {Moskalenko}\ and\ \citenamefont {Snoke}(2000)}]{Moskalenko_Snoke}%
  \BibitemOpen
  \bibfield  {author} {\bibinfo {author} {\bibfnamefont {S.~A.}\ \bibnamefont {Moskalenko}}\ and\ \bibinfo {author} {\bibfnamefont {D.~W.}\ \bibnamefont {Snoke}},\ }\href@noop {} {\emph {\bibinfo {title} {Bose-Einstein Condensation of Excitons and Biexcitons and Coherent Nonlinear Optics with Excitons}}}\ (\bibinfo  {publisher} {Cambridge University Press},\ \bibinfo {year} {2000})\BibitemShut {NoStop}%
\bibitem [{\citenamefont {Snoke}\ and\ \citenamefont {Keeling}(2017)}]{snoke2017new}%
  \BibitemOpen
  \bibfield  {author} {\bibinfo {author} {\bibfnamefont {D.~W.}\ \bibnamefont {Snoke}}\ and\ \bibinfo {author} {\bibfnamefont {J.}~\bibnamefont {Keeling}},\ }\bibfield  {title} {\bibinfo {title} {The new era of polariton condensates},\ }\href@noop {} {\bibfield  {journal} {\bibinfo  {journal} {Physics Today}\ }\textbf {\bibinfo {volume} {70}},\ \bibinfo {pages} {54} (\bibinfo {year} {2017})}\BibitemShut {NoStop}%
\bibitem [{\citenamefont {Littlewood}(2007)}]{littlewood2007condensates}%
  \BibitemOpen
  \bibfield  {author} {\bibinfo {author} {\bibfnamefont {P.}~\bibnamefont {Littlewood}},\ }\bibfield  {title} {\bibinfo {title} {Condensates made of light},\ }\href@noop {} {\bibfield  {journal} {\bibinfo  {journal} {Science}\ }\textbf {\bibinfo {volume} {316}},\ \bibinfo {pages} {989} (\bibinfo {year} {2007})}\BibitemShut {NoStop}%
\bibitem [{\citenamefont {Carusotto}\ and\ \citenamefont {Ciuti}(2013)}]{carusotto2013quantum}%
  \BibitemOpen
  \bibfield  {author} {\bibinfo {author} {\bibfnamefont {I.}~\bibnamefont {Carusotto}}\ and\ \bibinfo {author} {\bibfnamefont {C.}~\bibnamefont {Ciuti}},\ }\bibfield  {title} {\bibinfo {title} {Quantum fluids of light},\ }\href@noop {} {\bibfield  {journal} {\bibinfo  {journal} {Reviews of Modern Physics}\ }\textbf {\bibinfo {volume} {85}},\ \bibinfo {pages} {299} (\bibinfo {year} {2013})}\BibitemShut {NoStop}%
\bibitem [{\citenamefont {Peng}\ \emph {et~al.}(2022)\citenamefont {Peng}, \citenamefont {Tao}, \citenamefont {Haeberl{\'e}}, \citenamefont {Li}, \citenamefont {Jin}, \citenamefont {Fleming}, \citenamefont {K{\'e}na-Cohen}, \citenamefont {Zhang},\ and\ \citenamefont {Bao}}]{peng2022room}%
  \BibitemOpen
  \bibfield  {author} {\bibinfo {author} {\bibfnamefont {K.}~\bibnamefont {Peng}}, \bibinfo {author} {\bibfnamefont {R.}~\bibnamefont {Tao}}, \bibinfo {author} {\bibfnamefont {L.}~\bibnamefont {Haeberl{\'e}}}, \bibinfo {author} {\bibfnamefont {Q.}~\bibnamefont {Li}}, \bibinfo {author} {\bibfnamefont {D.}~\bibnamefont {Jin}}, \bibinfo {author} {\bibfnamefont {G.~R.}\ \bibnamefont {Fleming}}, \bibinfo {author} {\bibfnamefont {S.}~\bibnamefont {K{\'e}na-Cohen}}, \bibinfo {author} {\bibfnamefont {X.}~\bibnamefont {Zhang}},\ and\ \bibinfo {author} {\bibfnamefont {W.}~\bibnamefont {Bao}},\ }\bibfield  {title} {\bibinfo {title} {Room-temperature polariton quantum fluids in halide perovskites},\ }\href@noop {} {\bibfield  {journal} {\bibinfo  {journal} {Nature Communications}\ }\textbf {\bibinfo {volume} {13}},\ \bibinfo {pages} {7388} (\bibinfo {year} {2022})}\BibitemShut {NoStop}%
\bibitem [{\citenamefont {Snoke}\ \emph {et~al.}(2023)\citenamefont {Snoke}, \citenamefont {Hartwell}, \citenamefont {Beaumariage}, \citenamefont {Mukherjee}, \citenamefont {Yoon}, \citenamefont {Myers}, \citenamefont {Steger}, \citenamefont {Sun}, \citenamefont {Nelson},\ and\ \citenamefont {Pfeiffer}}]{snoke2023reanalysis}%
  \BibitemOpen
  \bibfield  {author} {\bibinfo {author} {\bibfnamefont {D.}~\bibnamefont {Snoke}}, \bibinfo {author} {\bibfnamefont {V.}~\bibnamefont {Hartwell}}, \bibinfo {author} {\bibfnamefont {J.}~\bibnamefont {Beaumariage}}, \bibinfo {author} {\bibfnamefont {S.}~\bibnamefont {Mukherjee}}, \bibinfo {author} {\bibfnamefont {Y.}~\bibnamefont {Yoon}}, \bibinfo {author} {\bibfnamefont {D.}~\bibnamefont {Myers}}, \bibinfo {author} {\bibfnamefont {M.}~\bibnamefont {Steger}}, \bibinfo {author} {\bibfnamefont {Z.}~\bibnamefont {Sun}}, \bibinfo {author} {\bibfnamefont {K.}~\bibnamefont {Nelson}},\ and\ \bibinfo {author} {\bibfnamefont {L.}~\bibnamefont {Pfeiffer}},\ }\bibfield  {title} {\bibinfo {title} {Reanalysis of experimental determinations of polariton-polariton interactions in microcavities},\ }\href@noop {} {\bibfield  {journal} {\bibinfo  {journal} {Physical Review B}\ }\textbf {\bibinfo {volume} {107}},\ \bibinfo {pages} {165302} (\bibinfo {year} {2023})}\BibitemShut {NoStop}%
\bibitem [{\citenamefont {Gevorkyan}\ and\ \citenamefont {Lozovik}(1985)}]{gevorkyan1985localization}%
  \BibitemOpen
  \bibfield  {author} {\bibinfo {author} {\bibfnamefont {Z.~S.}\ \bibnamefont {Gevorkyan}}\ and\ \bibinfo {author} {\bibfnamefont {Y.~E.}\ \bibnamefont {Lozovik}},\ }\bibfield  {title} {\bibinfo {title} {Localization and exciton spectrum in disordered systems},\ }\href@noop {} {\bibfield  {journal} {\bibinfo  {journal} {Sov. Phys. Solid State}\ }\textbf {\bibinfo {volume} {27}},\ \bibinfo {pages} {1079} (\bibinfo {year} {1985})}\BibitemShut {NoStop}%
\bibitem [{\citenamefont {Dolph}(1949)}]{dolph1949nonlinear}%
  \BibitemOpen
  \bibfield  {author} {\bibinfo {author} {\bibfnamefont {C.}~\bibnamefont {Dolph}},\ }\bibfield  {title} {\bibinfo {title} {Nonlinear integral equations of the hammerstein type},\ }\href@noop {} {\bibfield  {journal} {\bibinfo  {journal} {Transactions of the American Mathematical Society}\ }\textbf {\bibinfo {volume} {66}},\ \bibinfo {pages} {289} (\bibinfo {year} {1949})}\BibitemShut {NoStop}%
\bibitem [{\citenamefont {Kelley}(1995)}]{kelley1995iterative}%
  \BibitemOpen
  \bibfield  {author} {\bibinfo {author} {\bibfnamefont {C.~T.}\ \bibnamefont {Kelley}},\ }\href@noop {} {\emph {\bibinfo {title} {Iterative methods for linear and nonlinear equations}}}\ (\bibinfo  {publisher} {SIAM},\ \bibinfo {year} {1995})\BibitemShut {NoStop}%
\bibitem [{\citenamefont {Hopfield}(1958)}]{hopfield1958theory}%
  \BibitemOpen
  \bibfield  {author} {\bibinfo {author} {\bibfnamefont {J.}~\bibnamefont {Hopfield}},\ }\bibfield  {title} {\bibinfo {title} {Theory of the contribution of excitons to the complex dielectric constant of crystals},\ }\href@noop {} {\bibfield  {journal} {\bibinfo  {journal} {Physical Review}\ }\textbf {\bibinfo {volume} {112}},\ \bibinfo {pages} {1555} (\bibinfo {year} {1958})}\BibitemShut {NoStop}%
\bibitem [{\citenamefont {Fetter}\ and\ \citenamefont {Walecka}(2012)}]{fetter2012quantum}%
  \BibitemOpen
  \bibfield  {author} {\bibinfo {author} {\bibfnamefont {A.~L.}\ \bibnamefont {Fetter}}\ and\ \bibinfo {author} {\bibfnamefont {J.~D.}\ \bibnamefont {Walecka}},\ }\href@noop {} {\emph {\bibinfo {title} {Quantum theory of many-particle systems}}}\ (\bibinfo  {publisher} {Courier Corporation},\ \bibinfo {year} {2012})\BibitemShut {NoStop}%
\bibitem [{\citenamefont {Berman}\ \emph {et~al.}(2008)\citenamefont {Berman}, \citenamefont {Lozovik},\ and\ \citenamefont {Snoke}}]{berman2008theory}%
  \BibitemOpen
  \bibfield  {author} {\bibinfo {author} {\bibfnamefont {O.~L.}\ \bibnamefont {Berman}}, \bibinfo {author} {\bibfnamefont {Y.~E.}\ \bibnamefont {Lozovik}},\ and\ \bibinfo {author} {\bibfnamefont {D.~W.}\ \bibnamefont {Snoke}},\ }\bibfield  {title} {\bibinfo {title} {Theory of bose-einstein condensation and superfluidity of two-dimensional polaritons in an in-plane harmonic potential},\ }\href@noop {} {\bibfield  {journal} {\bibinfo  {journal} {Physical Review B—Condensed Matter and Materials Physics}\ }\textbf {\bibinfo {volume} {77}},\ \bibinfo {pages} {155317} (\bibinfo {year} {2008})}\BibitemShut {NoStop}%
\bibitem [{\citenamefont {Pathria}(2017)}]{pathria2017statistical}%
  \BibitemOpen
  \bibfield  {author} {\bibinfo {author} {\bibfnamefont {R.~K.}\ \bibnamefont {Pathria}},\ }\href@noop {} {\emph {\bibinfo {title} {Statistical Mechanics: International Series of Monographs in Natural Philosophy}}},\ Vol.~\bibinfo {volume} {45}\ (\bibinfo  {publisher} {Elsevier},\ \bibinfo {year} {2017})\BibitemShut {NoStop}%
\end{thebibliography}

\end{document}